\documentclass[
 reprint,
 amsmath,amssymb,
 aps,
prb,
floatfix,
]{revtex4-2}
\usepackage[english]{babel}
\usepackage{amsmath}
\usepackage{amssymb} 
\usepackage{amsthm}
\usepackage{bm}
\usepackage{braket}
\usepackage{dcolumn}
\usepackage{float}
\usepackage[bottom]{footmisc} 
\usepackage{graphicx}
\usepackage{hyperref}
\usepackage{lipsum}
\usepackage{physics}
\usepackage{placeins}
\usepackage{textcomp, mathcomp}
\usepackage{xcolor}
\usepackage{times}
\usepackage[normalem]{ulem} 

\newcommand{\supplementarysection}{%
  \setcounter{figure}{0}
  \let\oldthefigure\thefigure
  \renewcommand{\thefigure}{S\oldthefigure}
  \section*{Supplemental material}
 }

\begin{document}

\preprint{APS/123-QED}

\title{\Large Observation of the surface hybridization gap in the electrical transport properties of the ultrathin topological insulator (Bi$_{1-x}$Sb$_{x}$)$_2$Te$_3$}

\author{Feike van Veen\textsuperscript{1}} \thanks{These authors contributed equally to this paper.}
\author{Sofie Kölling\textsuperscript{1}}\thanks{These authors contributed equally to this paper.}
\author{Stijn R. de Wit\textsuperscript{1}}
\author{Roel Metsch\textsuperscript{1}}
\author{Daniel Rosenbach\textsuperscript{2}}
\author{Chuan Li\textsuperscript{1}}
\author{Alexander Brinkman\textsuperscript{1}}

\affiliation{\textsuperscript{1}MESA+ Institute for Nanotechnology, University of Twente, 7500 AE Enschede, The Netherlands}
\affiliation{\textsuperscript{2}II. Physikalisches Institut, Universität zu Köln, Zülpicher Strasse 77, D-50937 Köln, Germany}

\date{\today}
\begin{abstract}
We study the three-dimensional topological insulator (Bi\textsubscript{1-x}Sb\textsubscript{x})\textsubscript{2}Te\textsubscript{3} in its ultrathin limit i.e. when the thickness is of the same order as the surface state penetration depth. It is expected that in this limit a hybridization gap opens at the Dirac point, which gives rise to a quantum spin Hall (QSH) or insulating phase, depending on the material thickness. We fabricate (Bi\textsubscript{1-x}Sb\textsubscript{x})\textsubscript{2}Te\textsubscript{3} Hall bars with a thicknesses of 6 and 9 nm and measure an insulating phase around the Dirac point for low bias and at sub-Kelvin temperatures only in samples fabricated from the 6 nm films, which indicates the presence of a hybridization gap. The effect of a perpendicular magnetic field on the hybridization gap is studied but remains partially unresolved. The results form an important step towards experimentally realizing the quantum spin Hall state via hybridization in ultrathin films of (Bi\textsubscript{1-x}Sb\textsubscript{x})\textsubscript{2}Te\textsubscript{3}, yet, they also expose a knowledge gap regarding transport measurements in these systems.
\end{abstract}
\maketitle

\section{Introduction}
The surfaces of topological insulators (TIs) manifest massless Dirac states that are protected by time-reversal symmetry, while the bulk is electronically insulating \cite{fu2007topological}. Because of the exotic states at the interfaces, TIs have been of great interest to the scientific community \cite{hasan2010colloquium, kane2005Z2, bernevig_2006, Hsieh2008Topological}. In particular, two-dimensional TIs are predicted to exhibit dissipationless 1D modes at the edges, hence showcasing the quantum spin Hall (QSH) effect.

Signatures of the QSH effect are experimentally observed in a small number of materials. For instance, König et al. measured a conductance in two-dimensional (2D) HgTe/CdTe/HgTe quantum wells quantized to 2$e^2/h$ \cite{König2007quantum} with the electron charge $e$ and Planck's constant $h$, and a resistance close to $h/2e^2$ was found in monolayer WTe\textsubscript{2} by Wu et al. \cite{Wu2018observation}. Furthermore, bismuthene, i.e. a monolayer of bismuth, was presented as a conceptual platform for the QSH effect when stabilized by a SiC(0001) substrate \cite{Reis2017Bismuthene}. The aforementioned studies showed promising results, yet the fabrication of 2D materials poses considerable difficulties  \cite{kou2017two, Lunczer2019Approaching, konig2013spatially, Singh2019Lowenergy}. Therefore, it is of great interest to search for novel QSH candidates that circumvent these difficulties.

Such a notable candidate class is that of ultrathin three-dimensional topological insulators (3D TIs), where the thickness of the material is of the same order as the penetration depth of the surface states into the bulk of the material such that topologically protected states propagating on these surfaces overlap and hybridize \cite{zhang2010crossover, asmar2018topological}. Hybridization affects the density of states around the Dirac point (DP), and depending on film thickness can either lead to an insulating phase, or to a 1D QSH phase. 

Signatures of QSH states in ultrathin hybridized TIs have been observed by recent studies. For example, experiments using scanning tunneling spectroscopy have detected signatures of the QSH effect in both Bi\textsubscript{4}Br\textsubscript{4} and Bi\textsubscript{2}Se\textsubscript{3} \cite{shumiya2022evidence, moes2024characterization}, and earlier spectroscopy studies have detected a hybridization gap in Bi\textsubscript{2}Se\textsubscript{3} \cite{zhang2010crossover, neupane2014observation} and in Sb\textsubscript{2}Te\textsubscript{3} \cite{jiang2012landau}. Moreover, a spectroscopy study on thin films of BST showed signatures of a hybridization gap at 77 Kelvin \cite{Lang2012Competing}, yet,  a conclusive low temperature transport study of hybridized 3D TIs remains missing.

For this study, we deposit 6~nm thick (Bi\textsubscript{1-x}Sb\textsubscript{x})\textsubscript{2}Te\textsubscript{3}, with \textit{x} = 0.72 (hereafter referred to as BST) by molecular beam epitaxy (MBE) to probe the hybridization properties of a 3D TI in the ultrathin limit \cite{Mulder_Glind_2023}. BST is an alloy of two well-studied TIs Bi\textsubscript{2}Te\textsubscript{3} and Sb\textsubscript{2}Te\textsubscript{3} \cite{zhang2009topological, zhang2011band}. 
We focus on the experimental lower limit of film thickness that we have achieved in our MBE system, where the hybridization signatures should be maximal.

The conductance in the studied films is zero at low voltage bias and low temperature when gate-tuned close to the Dirac point. We exclude interface effects by performing four-terminal measurements and identify this gapped behavior as an intrinsic film property and attribute it to surface hybridization. Since the observed insulating phase for the studied film thickness and stoichiometry aligns with theoretical predictions, this study represents a significant step toward engineering quantum spin Hall devices using MBE-deposited BST.

\section{Ultrathin BST}
In the ultrathin limit of a 3D TI, the surface states hybridize when their penetration depth into the bulk of the material is of the same order as the thickness of the film $d$ \cite{zhang2010crossover}. Due to hybridization of the surface states a gap opens at the Dirac point. This can be captured in the Bernevig-Hughes-Zhang (BHZ) model \cite{bernevig_2006} as a tunnel coupling, $t(\mathbf{k}, d)$, between states on opposite surfaces\cite{asmar2018topological}, where $\mathbf{k}$ is the momentum vector. The term describing the decay of the \textit{z}-component of the surface states includes an imaginary exponent, which results in an oscillatory dependence of the hybridization on the film thickness. Depending on the sign of $t(\mathbf{k}, d)$, 1D edge states can still reside inside the bulk gap: a so-called positive gap implies an insulating state and a negative gap implies a QSH edge state.  \\

\begin{figure}
    \centering
    \includegraphics[width = \columnwidth]{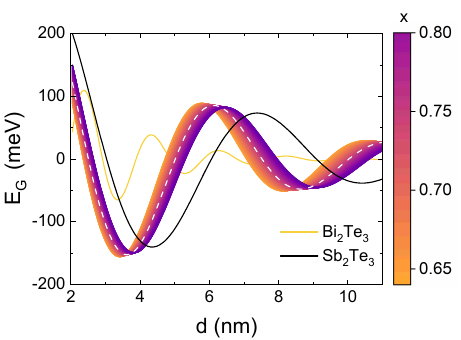}
    \caption{Magnitude of the hybridization gap, $E_\textup{G}$, as function of film thickness $d$ for varying ratios $x$ in (Bi\textsubscript{1-$x$}Sb\textsubscript{$x$})\textsubscript{2}Te\textsubscript{3}. Limiting cases $x = 0$ $(1)$ are plotted for the pure compounds Bi\textsubscript{2}Te\textsubscript{3}  (Sb\textsubscript{2}Te\textsubscript{3}) alongside the ratio $x=0.72$ (in white) used in the films studied in this work. }
    \label{fig:stijn}
\end{figure}

The topological phase of thin films therefore depends both on the thickness and on the material parameters (through $t(\mathbf{k}, d)$). For the pure constituent compounds of BST (Bi\textsubscript{2}Te\textsubscript{3} and Sb\textsubscript{2}Te\textsubscript{3}) the hybridisation gap $E_\textup{G} = 2t(\mathbf{k=0},d)$, shown in Fig.~\ref{fig:stijn} \cite{asmar2018topological}, oscillates with thickness -- but with different periods for the different compounds. In an alloy of (Bi\textsubscript{1-x}Sb\textsubscript{x})\textsubscript{2}Te\textsubscript{3}, one can interpolate \cite{teo_surface_2008,zhang2011band} the material parameters of Bi\textsubscript{2}Te\textsubscript{3} and Sb\textsubscript{2}Te\textsubscript{3} and compute the hybridization gap as a function of thickness for a range of stoichiometries $x$ (see Fig.~\ref{fig:stijn}). Based on this result, we expect an insulating phase for BST film thicknesses between 5-7 nm, and a nontrivial QSH phase between 3-5 and 8-10 nm. Note that the size of the gap decays exponentially with film thickness, indicating that signatures of hybridization are expected to be most pronounced for the thinnest films. Furthermore, in this low-energy model, the hybridization gap is expected to vary linearly with magnetic field \cite{supp}.

\section{Device Fabrication}
\label{sec:methods}
This work focuses on BST films with a stoichiometric ratio of $x=0.72$ and a thickness of 6 nm, the thinnest achievable closed layer using our MBE system, where hybridization effects are most prominent. 
The $x=0.72$ stoichiometry is chosen to ensure that the Dirac point is elevated from the bulk valence band with respect to pure Bi\textsubscript{2}Te\textsubscript{3} \cite{zhang2011band, zhang2009topological}. 

\begin{figure}
    \centering
    \includegraphics[width = \columnwidth]{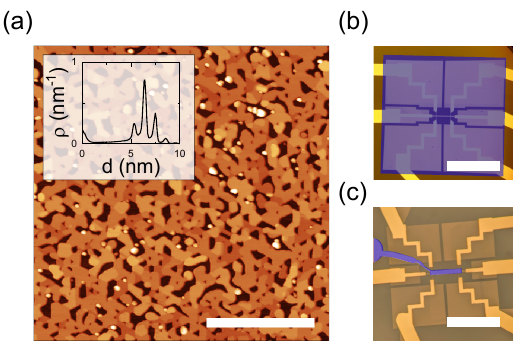}
    \caption{(a) Atomic force microscopy image of ultrathin BST (nominal thickness 6 nm) deposited on Al\textsubscript{2}O\textsubscript{3}. Holes in the film reach the substrate surface. The graph inset shows the height profile and the scale bar corresponds to 1 $\mu$m. This film is used to fabricate the device measured in Fig. \ref{fig:gateIV_Au}(a). (b) and (c) Optical microscopy picture of a generation 1 (Ti/Pd contacts) and generation 2 (Au contacts) BST Hall bar, respectively. The scale bars correspond to 100 $\mu$m.  The top gated area, covering both device and Ohmic contacts in generation 1 and only the region between voltage probes in generation 2 is shaded in blue. (b) was taken prior to depositing the Ti/Pd gate contact, so the gate metal itself is not shown.}
    \label{fig:hallbar}
\end{figure}

All samples were deposited using an Octoplus 300 MBE Systems from Dr. Eberl MBE Komponenten. 
Figure \ref{fig:hallbar}(a) contains an AFM scan of a representative BST film. The typical surface morphology is smooth within a few quintuple layers, and contains holes penetrating from the film surface to the substrate. An elaborate description of the thin film deposition can be found in an earlier work of Mulder \textit{et al.} \cite{Mulder_Glind_2023}.

Two device generations were fabricated with dissimilar metal contacts, of which the first generation was fabricated with Ti/Pd contacts. These contacts were evaporated onto the film and patterned using electron beam lithography (EBL) and lift-off. A thin layer ($\sim$5 nm) of titanium serves as an adhesion layer and is deposited between the BST film and the ($\sim$35 nm) palladium contact. Hall bars were patterned and etched in the BST thin films by means of EBL and argon beam etching. Subsequently, a thin layer ($\sim$30 nm) of AlO$_x$ was deposited by low-temperature atomic layer deposition (ALD) at $T = 100$ $^{\circ}$C to cap the device and to serve as a dielectric material. Lastly, Ti/Pd contacts were evaporated on top of the devices that were used as top gates. Since the ALD layer is deposited after the contacts, the insulating ALD layer allows the gate contact to cover the entire sample up to the Ohmic current and voltage probes. For this reason, the samples with Ti/Pd contacts are used among others to characterize the effect of gating on the BST thin film.

To exclude the opportunity that an observed gap is due to superconductivity, a second generation of devices is fabricated using gold contacts instead of Ti/Pd. A thin titanium layer can become superconducting at cryogenic temperatures \cite{steele1953superconductivity} which might, in combination with native oxide at the Ti/BST interface, form a superconducting-insulating-normal metal tunnel junction (SIN junction). In transport measurements SIN junctions are insulating when the superconducting gap exceeds the applied bias, and this insulating behavior could obscure a hybridization gap feature. Omitting the titanium adhesion layer was eliminated as a workaround, since then the superconducting material PdTe\textsubscript{2} might form on the interface between BST and the contact material \cite{bai2020novel}. 

The second generation of devices was fabricated in a different order: due to fabrication protocols, Au contacts were only sputter deposited onto the film after the gate dielectric was deposited with ALD. To remove the AlO$_x$  layer where the Au-contacts were to be sputtered, the AlO$_x$ was wet etched. Lastly, a Ti/Pd top gate contact is evaporated on the AlO$_x$-covered BST. To avoid shorting top gate and Ohmic contacts, these contacts are separated by a few microns (see Fig. \ref{fig:hallbar}), so in contrast with the first generation, the BST device is not gate-tunable as a whole. Although the device contains un-gated regions, the region probed in multiterminal Hall measurements is gated completely. The two device generations are shown in Fig. \ref{fig:hallbar}(b) and (c).  All measurements have been conducted in an Oxford Triton dilution refrigerator between 100 mK and 4.5 K.

\section{High T Characterization}\label{sec:normal}

\begin{figure}
    \centering
    \includegraphics[width=\columnwidth]{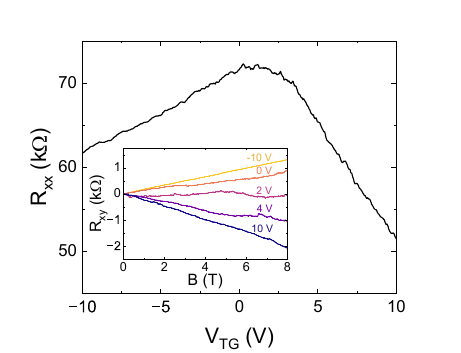}
    \caption{Longitudinal resistance R$_{xx}$ as function of top gate voltage at $B = 0\;$T. Inset: Anti-symmetrized Hall resistance versus out-of-plane magnetic field for varying top gate voltages, as indicated next to the graphs. The data was obtained on a $L \cross W = 14.8\; \mu \mathrm{m} \cross 3.2 \;\mu \mathrm{m}$ Hall bar with Ti/Pd contacts in a 4 terminal configuration at $T = 4.5$ K. This data was obtained using standard lock-in techniques.}
    \label{fig:RxxRxy}
\end{figure}

At $T = 4.5$ K, we characterized intrinsic electronic material properties using a Hall bar configuration similar to the device shown in Fig. \ref{fig:hallbar}. The longitudinal resistance versus top gate voltage ($V_\mathrm{TG}$) at $B = 0$ T is depicted in Fig. \ref{fig:RxxRxy}, which shows a maximum around $V_\mathrm{TG}$ = 2 V. We find that the chemical potential is at the Dirac point for $V_\mathrm{TG}$ = 2 V and demonstrate that it moves through the upper and lower segment of the Dirac cone for higher and lower gate voltages respectively \footnote{Although a Dirac cone is theoretically described by a linear dispersion, in reality, the upper and lower segment of the cone are not symmetric around the DP \cite{zhang2009topological}, which is reflected by the difference in slope left and right from $V_\mathrm{TG}$ = 2 V.}. This is supported by the measured Hall resistances at several top gate voltages as shown in the inset of Fig. \ref{fig:RxxRxy}. Here, for $V_\mathrm{TG}$ = -10 V an $V_\mathrm{TG}$ = 10 V the slope $\dd R_{xy}/\dd B$ is evidently opposite in sign. This indicates that a carrier inversion has taken place when moving the gate voltage between these two extremes.

Furthermore, assuming that the transport for $V_\mathrm{TG}$ = -10 V and $V_\mathrm{TG}$ = 10 V is dominated by a single carrier type, the Hall resistances of these curves correspond to carrier densities $n_{h}$ = 3.8$\times 10^{12}$ cm$^{-2}$ and $n_{e}$ = 2.6$\times 10^{12}$ cm$^{-2}$ respectively. Here we make use of the inverse relationship between the slope $\partial R_{xy}/\partial B$ and the carrier density in a single-band Drude model. We argue that the curves corresponding to $V_{\mathrm{TG}}$ = 0 V and $V_{\mathrm{TG}}$ = 4 V do not correspond to a higher carrier density than those of $V_{\mathrm{TG}}$ = 10 V and $V_{\mathrm{TG}}$ = 10 V, and instead, the slopes are lower due to carrier mixing close to the charge neutrality point (CNP) \cite{he2012highly, li2017interaction, zhang2011growth, chen2011tunable}. An illustrative example is the (almost) flat Hall resistance found for $V_{\mathrm{TG}}$ = 2 V, which we interpret as the chemical potential coinciding with the CNP, i.e. the DP in topological insulators.

Lastly, assuming linear dispersion $E_F = \hbar v_F k_F$, with $v_F = 4.7\cdot 10^5$ m/s \cite{mulder2022spectroscopic}, the carrier densities correspond to an energy window between -150 meV and 125 meV, which falls within the typical band gap in BST \cite{zhang2011band}. All in all, we argue that we can successfully tune the chemical potential through the DP and that for this sample the states around DP are probed for a gate voltage close to 2V. \newline 

\section{Observation of a hybridization gap}\label{sec:gapped}
\begin{figure*}
    \centering
    \includegraphics{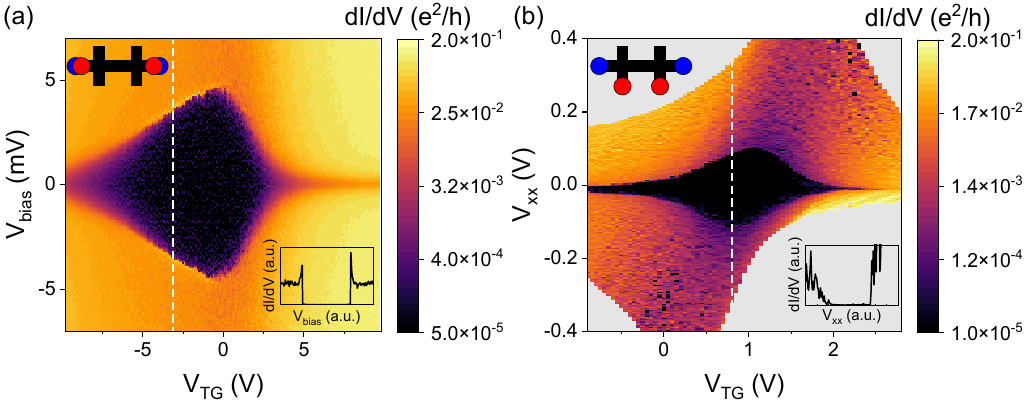}
    \caption{Differential conductance ($\dd I / \dd V$) as function of $V_\mathrm{TG}$ and (a) $V_\mathrm{bias}$ or (b) $V_{xx}$ at $T = 100$ mK. Blue dots on the inset Hall bars denote the current path and red dots denote the voltage probes. (a) 2-terminal data obtained on a $L \times W = 148$ $\mu$m$\times 25.6$ $\mu$m Hall bar with Ti/Pd contacts. (b) 4-terminal data obtained on a $L \times W = 80$ $\mu$m$\times 5$ $\mathrm{\mu}$m Hall bar with Au contacts. The insets in the bottom right of (a) and (b) show linecuts of the two figures respectively. The position of each of the linecuts is indicated by a white dotted line.}
    \label{fig:gateIV_Au}
\end{figure*}
\begin{figure}
    \centering
    \includegraphics{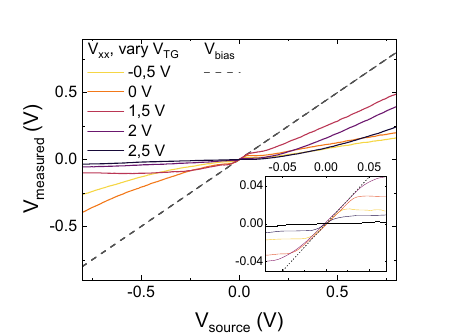}
    \caption{Comparison between 2-terminal ($V_\mathrm{bias}$) and 4-terminal ($V_{xx}$) voltage as a function of source voltage bias. The 2-terminal voltage ($V_\mathrm{bias}$) follows the source voltage (we use a current measurement to convert these traces to $\dd I/\dd V$). The 4-terminal voltage shows a slope change when the gapped region ($I = 0$) is entered. Here, $V_\mathrm{bias}=V_{xx}$ (see inset): the entire bias voltage drops over the region between the voltage probes. }
    \label{fig:biasV_Vxx}
\end{figure}
In this section, we search for signatures of a hybridization gap at low temperature ($T = 100$ mK). For that purpose, we apply a voltage bias to the ultrathin TI devices and measure the current ($I$), while varying bias ($V_\mathrm{bias}$) and $V_\mathrm{TG}$. We discuss both two-terminal and four-terminal measurements \footnote{Due to a lower fabrication yield of the generation 1 devices, it was only possible to use a 2-terminal setup.}: in two-terminal measurements, the differential conductance ($\dd I/\dd V_\mathrm{bias}$) includes contact resistance effects. These effects are excluded in the four-terminal measurements, where the differential conductance is calculated from the  voltage drop between the Hall bar probes ($\dd I/\dd V_{xx}$). As mentioned in section \ref{sec:methods}, we fabricated two generations of devices, with Ti/Pd (generation 1) or Au (generation 2) contacts. 
These devices are fabricated from different BST films, albeit with the same deposition parameters as the sample in Fig. \ref{fig:RxxRxy}.

The data in Fig. \ref{fig:gateIV_Au}(a) is measured on a generation 1 BST device, similar to the device in Fig. \ref{fig:RxxRxy}, and most strikingly, it shows that an insulating gap opens around the Dirac point, as the conductivity is reduced to a value smaller than a permille of $e^2/h$. Clearly, the sample becomes conducting when either the chemical potential is tuned away from the DP, or when an increased voltage bias is applied. Furthermore, for $-5$ V$<V_\mathrm{TG}<0$ V, the differential conductance is enhanced at the edge of the gap. The enhanced conductivity could be due to the enhanced density of states of a superconductor just above the gap edge. 

To exclude this possibility we perform a control experiment on a device from a similar film contacted using sputtered Au (generation 2) rather than Ti/Pd. Generation 2 devices allow for 4-terminal measurements, whereby contact resistance is excluded. Also here, we observe a gapped spectrum, as can be seen in Fig. \ref{fig:gateIV_Au}(b). In contrast to Fig.\ref{fig:gateIV_Au}(a), the enhanced differential conductance at the edge of the gap has disappeared, which can be attributed to the absence of any superconducting material. The map is skewed due to the interplay between the bias and gate voltage, since both are of similar order \cite{kolling2025gate}. 

In the gapped region, we observe that $V_{xx}$ equals $V_\mathrm{bias}$: the bias voltage drops entirely over the (top-gated) region between the voltage probes as visible in Fig. \ref{fig:biasV_Vxx}. If the gap would be induced by a (Schottky) barrier effect at the metal/TI interface, we would expect that the voltage drop is located at the source/drain contacts, and not at the central region of the device. Therefore we conclude that the insulating character stems from the ultrathin film itself, and not from an interface/contact effect. Based on the predictions from theory, this insulating phase can be caused by surface hybridization.

\begin{figure}[h!]
    \centering
    \includegraphics[width = \linewidth]{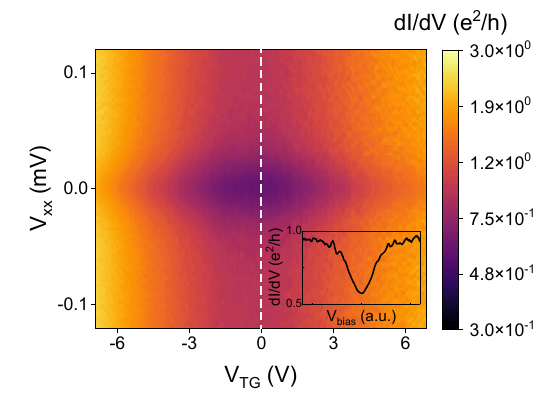}
    \caption{Differential conductance ($\dd I/\dd V_{xx}$) of a BST Hall bar device with thickness 9 nm, as a function of applied gate voltage ($V_\mathrm{TG}$) and longitudinal voltage ($V_\mathrm{xx}$). The measurement is performed with a current source, however, for comparison with Fig. \ref{fig:biasV_Vxx}, the data is plotted as function of longitudinal voltage. The data was obtained in a 4-terminal configuration on a $L \times W$ = 6.125 $\mathrm{\mu m}$ $\times$ 1.25 $\mathrm{\mu m}$ Hall bar with Ti/Pd contacts at $T = 100$ mK.}
    \label{fig:suppl_9nm}
\end{figure}

Moreover, we have performed an identical measurement, albeit with a current source, on a BST film with a thickness of 9 nm, which is shown in Fig. \ref{fig:suppl_9nm}. Here, a slight reduction in conductivity is observed for similar gate voltages where a gap was found in the 6 nm films. Although the reduction in conductivity is more pronounced for low bias voltages, the strong insulating behavior is absent. While the topological nature of the gap in the 6 nm film corresponds to the model described in Fig. \ref{fig:stijn}, the absence of a quantized conductance in the 9 nm film contradicts with this model, which predicts a nontrivial QSH phase for this thickness.\\

The absence of a QSH phase in the 9 nm film can have multiple reasons. Firstly, the expected gap for a 9 nm thin film is significantly smaller than for a film of 6 nm. Therefore, observation of the gap in 9 nm thin film can be obscured by fluctuations in the potential landscape that are of similar order to the gap size \cite{Huang2022Conductivity}. Secondly, it is found in other systems that QSH are much less robust against disorder then predicted, hypothetically leading to a lower conductance than the expected $e^2/h$ per channel \cite{Lunczer2019Approaching}. Apart from surface hybridization, a reduction of the conductance as visible in Fig. \ref{fig:suppl_9nm} could also arise from surface-state conduction, in which electron-electron interactions are being suppressed via Joule heating at elevated bias \cite{kolling2025non, anderson1979possible}. The reduction in conductance by itself is therefore not a sufficient proof of hybridization.

Notably, recent on work surface hybridization in the BHZ model describes the partial invalidity of the four-band model used for this study \cite{maisel2024topology}. In particular, it was found that by considering only the four bands that are lowest in energy, the topological character of the hybridization gap in Bi\textsubscript{2}Se\textsubscript{3} can be found to be trivial for a thickness of 4 and 5 nm, while including the lowest eight bands yields a non-trivial character, which corresponds to experimental findings \cite{moes2024characterization}. For Bi\textsubscript{2}Se\textsubscript{3} films with thickness $d>6$ nm even the eight-band model is insufficient for determining the (topological character of the) hybridization gap. Additionally, due to surface roughness, not all of the film has the same thickness. Therefore, we stress that the four-band model used here should only be considered as crude estimate of the hybridization gap in BST thin films.

\section{Influence of Magnetic Field}
\begin{figure*}
    \centering
    \includegraphics[width=\textwidth]{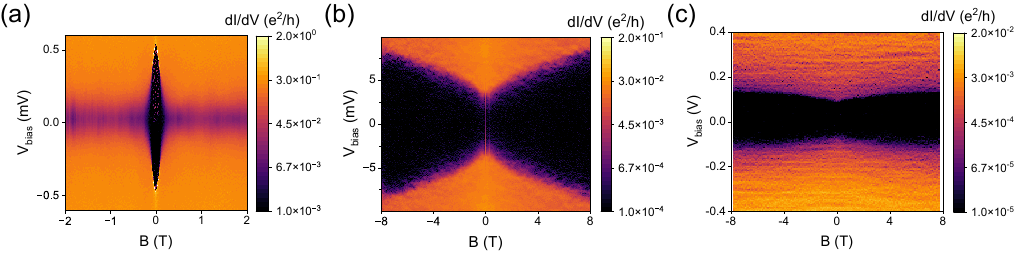}
    \caption{$\dd I/ \dd V$ as a function of $V_\mathrm{bias}$ and magnetic field, highlighting a varying magnetic field response of the gap. Each map is obtained on a device from a different film, (a) and (b) are obtained with Ti/Pd contacts, and (c) with Au contacts ($L = 37.5$ $\mu$m). (a) A gap due to superconductivity closes at 100 mT. Outside this gap, the $\dd I/ \dd V$ is nonzero for all bias voltages. (b) Around zero, the gap slightly reduces for finite fields due to superconductivity. For larger fields, the gap increases. (c) For increasing magnetic field, the gap increases over the entire range between 0 and 8 T.}
    \label{fig:gapvsB_devicecomp}
\end{figure*}

In this section, we study the dependence of the hybridization gap on an out-of-plane applied magnetic field. In Fig. \ref{fig:gapvsB_devicecomp}, we tune the top gate voltage to the Dirac point and measure IV curves as function of magnetic field. The 2-terminal results for two different Ti/Pd-contacted devices are shown in Fig. \ref{fig:gapvsB_devicecomp}(a) and (b), and for the  Au-contacted device in Fig. \ref{fig:gapvsB_devicecomp}(c) ($L \times W = 37.5$ $\mu$m$\times 5$ $\mu$m). In both Ti/Pd devices, a small reduction of the gap for $B< 200$ mT is visible. Moreover, only below this crossover value the coherence peaks are visible. The low-field suppression of the gap is absent in the Au-contacted device. We attribute this low-field behavior to a suppression of a superconducting gap.

The scaling of the gap for $B>200$ mT varies from device to device. In Fig. \ref{fig:gapvsB_devicecomp}(a), the gap is closed at higher magnetic fields ($\dd I/ \dd V$ stays nonzero), whereas in Fig. \ref{fig:gapvsB_devicecomp}(b) the gap increases at higher magnetic fields. A similar behavior is observed for the Au-contacted device in Fig. \ref{fig:gapvsB_devicecomp}(c).

\begin{figure}
    \centering
    \includegraphics[width = \linewidth]{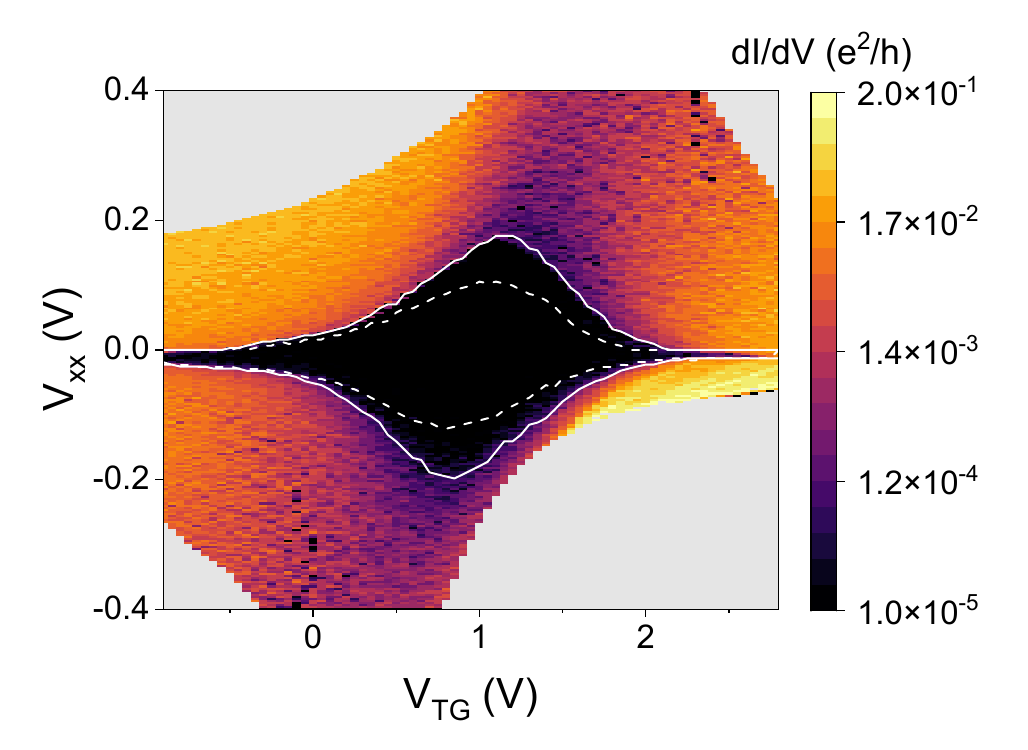}
    \caption{$\dd I/ \dd V$ at $B = 8$ T as a function of $V_\mathrm{TG}$ and $V_{xx}$ in a 4-terminal setup on the Au-contacted device from Fig. 4(b) in the main text ($L = 80$ $\mu$m). The solid and dashed lines correspond to the extracted gap at $B = 8$ T and $B = 0$ T, respectively. For the $B = 0$ T dataset, see Fig. 4(b).}
    \label{fig:4term_gap8T}
\end{figure}

We compare the gap at 0 T and 8 T for the Au-contacted, $L \times W = 80$ $\mu$m$\times 5$ $\mu$m, 4-terminal device in Fig. \ref{fig:4term_gap8T}. Similar to the device in Fig. \ref{fig:gapvsB_devicecomp}(c), the gap at 8 T is larger than at 0 T. The increase is largest around the Dirac point, hardly any increase is observed for $V_\mathrm{TG}<0$ and $V_\mathrm{TG}>1.5$.

The opening of a band gap with increased magnetic field is in line with earlier measurements and theory on similar Bi-based compounds \cite{fatemi2014electrostatic} and it can be expected based on the entrance of a Zeeman gap in addition to the intrinsic hybridization \cite{supp} gap. In the latter description, however, the measured gap is predicted to scale linearly with $B$ by a factor of $g \mu_B$, whereas the data in Fig.~\ref{fig:gapvsB_devicecomp}(b) and (c) suggest a nonlinear dependence.
Furthermore, we do not observe a crossover between an insulating state and a chiral state with quantized conductance as predicted by some models \cite{wang2015quantum}, which might indicate that further theoretical work on hybridization in ultrathin BST with a disordered potential landscape is required.

As mentioned in the previous section, the gap extracted from IV curves is not expected to correspond directly to the hybridization gap due to a disordered potential landscape in our ultrathin films \cite{nandi2018signatures}. However, if the (geometry-independent) temperature dependence is an indication of hybridization gap size, we can estimate at which magnetic fields the Zeeman gap would exceed the hybridization gap.
By performing temperature-dependent measurements we find that the gap closes at $T \approx 3$ K \cite{refsupp}\nocite{hikami1980spin, steinberg2011electrically, lu2011competition, Lu2011Weak, chong2019tunable, yang2015dual, Skinner2013Theory, knispel2017charge, Mahmoodian2020Conductivity}, which would imply $E_\mathrm{gap} \approx 0.26$ meV. Then, in the low energy regime (see \cite{supp}) $\Delta \sim E_\mathrm{gap}$ would require $B = \Delta/g\mu_B \sim 10^{-1}$ T (using $g = 10-30$ \cite{drath1967band}). This value falls well within the range over which we measured a continuous increase of the gap in Fig. \ref{fig:gapvsB_devicecomp}(c). Thus, a Zeeman gap does not directly add to the intrinsic gap observed in BST and the effects of a magnetic field on the gap require further research.

\section{Conclusion}
We experimentally observed an 
insulating phase in 6 nm thin films of (Bi\textsubscript{0.28}Sb\textsubscript{0.72})\textsubscript{2}Te\textsubscript{3} deposited by MBE. This insulating phase is measured consistently in various devices fabricated from a BST film of thickness 6 nm, using either Ti/Pd or Au contacts, and in each device, the gap is found to close by tuning the gate voltage or applying a sufficiently large bias. The gap size is found to be the largest close to the intrinsic position of the chemical potential, which is tuned by controlling the Bi:Sb ratio. Samples with Ti/Pd contacts are used to maximize the gate-tunable area of the device, while samples with Au contacts are used to exclude the role of a possible superconducting interface in our devices, showing that insulating phase is intrinsic to the TI. Upon measuring a BST device fabricated from a film with thickness 9 nm, the insulating phase vanishes. Combining these considerations, we attribute the insulating phase to a hybridization effect.

In a small number of devices (see \cite{refsupp}) the gap size is much smaller than the theoretically predicted gap, which can be attributed to the presence of charge puddles \cite{Huang2022Conductivity}, however, for other samples the measured gap is much larger than expected. We would like to suggest that large insulating samples cause the applied bias potential to smear out over the probed sample. This effect thus introduces a geometry dependence of the applied bias voltage required to overcome the insulating characteristics. To elucidate this suggestion further a discussion is added to the Supplemental Material \cite{refsupp}. Furthermore, the potential landscape might be such that a large bias does not inject carriers over the hybridization gap but instead allows the carriers to tunnel from one impurity to the next. Hence, instead of the hybridization gap, the impurity density is the factor that determines the measurable gap.

An important observation is that the effect of a magnetic field, showing large sample-to-sample variations, remains an open question. In all cases, our results contradict predicted transitions from a trivially insulating regime to a chiral quantized regime, and neither do we observe the onset of a linear gap closure and reopening due to a Zeeman shift. Therefore, further research is required, consisting of reevaluating the current theoretical frameworks on hybridization effects, and considering whether other thickness-dependent effects could give rise to a similar insulating phase in ultrathin topological insulators.

We believe that our findings contribute to further understanding of surface hybridization in topological insulators. Specifically, we showcase the potential of MBE-deposited ultrathin films by which the thickness can be neatly controlled, with an intrinsic chemical potential close to the Dirac point. This study opens up the possibility of studying the hybridization gap as function of film thickness, hereby experimentally examining the plausible presence of an oscillatory dependence of the gap on the film thickness. Ultimately, these studies could lead to measuring the QSH effect in scalable and controllable platforms.

\begin{acknowledgments}
We would like to thank Frank Roesthuis and Daan Wielens for their support and fruitful discussions. Moreover, we acknowledge the previous work of Liesbeth Mulder for laying the groundwork for this study.

This research was financially supported by a Lockheed Martin Corporation Research Grant and the research program “Materials for the Quantum Age” (QuMat). This program (registration number 024.005.006) is part of the Gravitation program financed by the Dutch Ministry of Education, Culture and Science (OCW).
\end{acknowledgments}

\FloatBarrier
\bibliography{references}

@article{Hsieh2008Topological,
    author = {D. Hsieh and D. Qian and L. Wray and Y. Xia and Y. S. Hor and R. J. Cava and M. Z. Hasan},
    title = {A topological Dirac insulator in a quantum spin {Hall} phase},
    journal = {Nature},
    year = {2008},
    volume = {452},
    pages = {970-974}
}

@article{Reis2017Bismuthene,
    author = {F. Reis and  G. Li and L. Dudy and M. Bauernfeind and S. Glass and W. Hanke and R. Thomale and J. Schäfer and R. Claessen},
    title = {Bismuthene on a SiC substrate: A candidate for a high-temperature quantum spin {Hall} material},
    journal = {Science},
    year = {2017},
    volume = {357},
    pages = {287-290},
}

@article{Singh2019Lowenergy,
    author = {Sobhit Singh and Zeila Zanolli and Maximilian Amsler and Brahim Belhadji and Jorge O. Sofo and Matthieu J. Verstraete and  Aldo H. Romero},
    title = {Low-Energy Phases of {Bi} Monolayer Predicted by Structure Search in Two Dimensions},
    journal = {The Journal of Physical Chemistry Letters},
    year = {2019},
    volume = {10},
    pages = {7324-7332},
}

@article{knispel2017charge,
  title={Charge puddles in the bulk and on the surface of the topological insulator {BiSbTeSe$_2$} studied by scanning tunneling microscopy and optical spectroscopy},
  author={Knispel, T and Jolie, W and Borgwardt, N and Lux, J and Wang, Zhiwei and Ando, Yoichi and Rosch, A and Michely, T and Gr{\"u}ninger, M},
  journal={Physical Review B},
  volume={96},
  number={19},
  pages={195135},
  year={2017},
}

@article{teo_surface_2008,
    title = {Surface States of the Topological Insulator {Bi$_{1-x}$Sb$_{x}$}},
    volume = {78},
    issn = {1098-0121, 1550-235X},
    number = {4},
    journal = {Physical Review B},
    author = {Teo, Jeffrey C. Y. and Fu, Liang and Kane, C. L.},
    month = jul,
    year = {2008},
    pages = {045426},
}

@article{bernevig_2006,
    author = {B. Andrei Bernevig  and Taylor L. Hughes  and Shou-Cheng Zhang },
    title = {Quantum Spin {Hall} Effect and Topological Phase Transition in {HgTe} Quantum Wells},
    journal = {Science},
    volume = {314},
    number = {5806},
    pages = {1757-1761},
    year = {2006},
}

@article{Mulder_Glind_2023,
    author={Mulder, L. and van de Glind, H. and Brinkman, A. and Concepción, O.},
    year={2023},
    title={Enhancement of the Surface Morphology of {(Bi\textsubscript{0.4}Sb\textsubscript{0.6})\textsubscript{2}Te\textsubscript{3}} Thin Films by In Situ Thermal Annealing},
    journal={Nanomaterials},
    volume={13},
    number={762},
}

@article{zhang2009topological,
  title={Topological insulators in {Bi$_{2}$Se$_{3}$}, {Bi$_{2}$Te$_{3}$} and {Sb$_{2}$Te$_{3}$} with a single Dirac cone on the surface},
  author={Zhang, Haijun and Liu, Chao-Xing and Qi, Xiao-Liang and Dai, Xi and Fang, Zhong and Zhang, Shou-Cheng},
  journal={Nature Physics},
  volume={5},
  number={6},
  pages={438-442},
  year={2009},
}

@article{zhang2011band,
  title={Band structure engineering in {(Bi$_{1-x}$Sb$_x$)$_2$Te$_3$} ternary topological insulators},
  author={Zhang, Jinsong and Chang, Cui-Zu and Zhang, Zuocheng and Wen, Jing and Feng, Xiao and Li, Kang and Liu, Minhao and He, Ke and Wang, Lili and Chen, Xi and Xue, Qi-Kun and Ma, Xucun and Wang, Yuyu},
  journal={Nature Communications},
  volume={2},
  number={1},
  pages={574},
  year={2011},
  publisher={Nature Publishing Group UK London}
}

@article{hikami1980spin,
    author = {Hikami, S. and Larkin, A. I. and Nagaoka, Y.},
    title = {Spin-orbit interaction and magnetoresistance in the two dimensional random system},
    journal = {Progress of Theoretical Physics},
    volume = {63},
    number = {2},
    pages = {707-710},
    year = {1980},
}

@article{asmar2018topological,
    author = {Asmar, Mahmoud. M. and Sheehy, Daniel. E. and Vekhter, Ilya},
    title = {Topological phases of topological-insulator thin films},
    journal = {Physical review B},
    volume = {97},
    number = {7},
    pages = {075419},
    year = {2018},
}

@article{König2007quantum,
    author = {König, M. and Wiemann, S. and Brüne, C. and Roth, A. and Buhmann, H. and Molenkamp, L.W. and Qi, X.-L. and Zhang, S.-C.},
    title = {Quantum spin {Hall} insulator state in {HgTe} qauntum wells},
    journal = {Science},
    year = {2007},
    volume = {318},
    number = {3851},
    pages = {799-770},
}

@article{Wu2018observation,
    author = {Wu, Sanfeng and Fatemi, Valla and Gibson, Quinn D. and Watanabe, Kenji and Taniguchi, Takashi and Cava, Robert J.},
    title = {Observation of the quantum spin {Hall} effect up to 100 {Kelvin} in a monolayer crystal},
    journal = {Science},
    year = {2018},
    volume = {359},
    number = {6371},
    pages = {76-79},
}

@article{Lu2011Weak,
    author = {Lu, Hai-Zhou and Shen, Shun-Qing},
    title = {Weak localization of bulk channels in topological insulator thin films},
    journal = {Physical Review B},
    year = {2011},
    volume = {84},
    number = {12},
    pages = {125138},
}

@article{Lang2012Competing,
    author = {Lang, Murong and He, Liang and Kou, Xufeng and Upadhayaya, Pramey and Fan, Yabin and Chu, Hao and Jiang, Ying and Bardarson, Jens H. and Jiang, Wanjun and Choi, Eun Sang and Wang, Yong and Yeh, Nai-Chang and Moore, Joel and Wang, Kang L.},
    title = {Competing weak localization and weak antilocalization in ultrathin topological insulators},
    journal = {Nano Letters},
    year = {2012},
    volume = {13},
    number = {1},
    pages = {48-53},
}

@article{steinberg2011electrically,
  title={Electrically tunable surface-to-bulk coherent coupling in topological insulator thin films},
  author={Steinberg, Hadar and Lalo{\"e}, J-B and Fatemi, Valla and Moodera, Jagadeesh S and Jarillo-Herrero, Pablo},
  journal={Physical Review B—Condensed Matter and Materials Physics},
  volume={84},
  number={23},
  pages={233101},
  year={2011},
  publisher={APS}
}

@article{Lunczer2019Approaching,
    author = {Lunczer, Lukas and Leubner, Philipp and Endres, Martin and Müller, Valentin L. and Brüne, Christoph and Buhmann, Hartmut and Molenkamp, Laurens W.},
    title = {Approaching quantization in macroscopic quantum spin {Hall} devices through gate training},
    journal = {Physical Review Letters},
    year = {2019},
    volume = {123},
    number = {4},
    pages = {047701},
}

@article{mulder2022spectroscopic,
  title={Spectroscopic signature of surface states and bunching of bulk subbands in topological insulator {(Bi$_{0.4}$Sb$_{0.6}$)$_2$Te$_3$} thin films},
  author={Mulder, Liesbeth and Castenmiller, Carolien and Witmans, Femke J and Smit, Steef and Golden, Mark S and Zandvliet, Harold JW and de Boeij, Paul L and Brinkman, Alexander},
  journal={Physical Review B},
  volume={105},
  number={3},
  pages={035122},
  year={2022},
  publisher={APS}
}

@article{Huang2022conductivity,
  title={Conductivity of two-dimensional small gap semiconductors and topological insulators in strong {Coulomb} disorder},
  author={Huang, Yi and Skinner, Brian and Shklovskii, BI},
  journal={Journal of Experimental and Theoretical Physics},
  volume={135},
  number={4},
  pages={409--425},
  year={2022},
  publisher={Springer}
}

@article{nandi2018signatures,
  title={Signatures of long-range-correlated disorder in the magnetotransport of ultrathin topological insulators},
  author={Nandi, D and Skinner, B and Lee, GH and Huang, K-F and Shain, K and Chang, Cui-Zu and Ou, Y and Lee, S-P and Ward, J and Moodera, JS and Kim, P and Halperin, BI and Yacoby, A},
  journal={Physical Review B},
  volume={98},
  number={21},
  pages={214203},
  year={2018},
  publisher={APS}
}

@article{shumiya2022evidence,
    title={Evidence of a room-temperature quantum spin {Hall} edge state in a higher-order topological insulator},
    author={Nana Shumiya and Md Shafayat Hossain and Jia-Xin Yin and Zhiwei Wang and Maksim Litskevich and Chiho Yoon and Yongkai Li and Ying Yang and Yu-Xiao Jiang and Guangming Cheng and Yen-Chuan Lin and Qi Zhang and Zi-Jia Cheng and Tyler A. Cochran and Daniel Multer and Xian P. Yang and Brian Casas and Tay-Rong Chang and Titus Neupert and Zhujun Yuan and Shuang Jia and Hsin Lin and Nan Ya and Luis Balicas and Fan Zhang and Yugui Yao and M. Zahid Hasan},
    journal={Nature Materials},
    volume={21},
    number={10},
    pages={1111-1115},
    year={2022},
    publisher={Springer}
}

@article{moes2024characterization,
    author = {Jesper R. Moes and Jara F. Vliem and Pedro M. M. C. de Melo and Thomas C. Wigmans and Andrés R. Botello-Méndez and Rafael G. Mendes and Ella F. van Brenk and Ingmar Swart and Lucas Maisel Licerán and Henk T.C. Stoof and Christophe Delerue and Zeila Zanolli and Daniel Vanmaekelbergh},
    title = {Characterization of the Edge States in Colloidal {Bi$_2$Se$_3$} Platelets},
    journal = {Nano Letters},
    volume = {24}, 
    number = {17},
    pages = {5110-5116},
    year = {2024},
    publisher = {Springer}
}

@article{maisel2024topology,
      title={Topology of {Bi$_2$Se$_3$} nanosheets},
      author={Maisel Licer{\'a}n, Lucas and Koerhuis, SJH and Vanmaekelbergh, Daniel and Stoof, HTC},
      journal={Physical Review B},
      volume={109},
      number={19},
      pages={195407},
      year={2024},
      publisher={APS}
}

@article{hasan2010colloquium,
    author = {Hasan, M. Z. and Kane, C. L.},
    title = {Colloquium: Topological insulators},
    journal = {Reviews of Modern Physics},
    volume = {82}, 
    number = {4},
    pages = {3045-3067},
    year = {2010},
    publisher = {APS}
}

@article{kane2005Z2,
    author = {C. L. Kane and E. J. Mele},
    title = {${Z}_{2}$ Topological Order and the Quantum Spin {Hall} Effect},
    journal = {Physics Review Letters},
    volume = {95},
    number = {14},
    pages = {146802},
    year = {2005},
    publisher = {APS}
}

@article{fu2007topological,
    author = {Fu, L. and Kane, C. L. and Mele, E. J.},
    title = {Topological Insulators in Three Dimensions},
    journal = {Physics Review Letters},
    year = {2007},
    volume = {98},
    number = {10},
    pages = {106803},
    publisher = {APS}
}

@article{zhang2010crossover,
    author = {Yi Zhang and Ke He and Cui-Zu Chang and Can-Li Song and Li-Li Wang and Xi Chen and Jin-Feng Jia and Zhong Fang and Xi Dai and Wen-Yu Shan and Shun-Qing Shen and Qian Niu and Xiao-Liang Qi and Shou-Cheng Zhang and Xu-Cun Ma and Qi-Kun Xue},
    title = {Crossover of the three-dimensional topological insulator {Bi$_2$Se$_3$} to the two-dimensional limit},
    journal = {Nature Physics},
    year = {2010},
    volume = {6},
    number = {8},
    pages = {584-588},
    publisher = {Springer}
}

@article{bai2020novel,
    author = {Mengmeng Bai and Fan Yang and Martina Luysberg and Junya Feng and Andrea Bliesener and Gertjan Lippertz and A. A. Taskin and Joachim Mayer and Yoichi Ando},
    title = {Novel self-epitaxy for inducing superconductivity in the topological insulator {(Bi$_{1-x}$Sb$_x$)$_2$Te$_3$}},
    journal = {Physical Review Materials},
    year = {2020},
    volume = {4},
    number = {9},
    pages = {094801},
    publisher = {APS},
}

@article{steele1953superconductivity,
  title={Superconductivity of titanium},
  author={Steele, MC and Hein, RA},
  journal={Physical Review},
  volume={92},
  number={2},
  pages={243},
  year={1953},
  publisher={APS}
}

@article{wang2015quantum,
  title={Quantum anomalous {Hall} effect in magnetic topological insulators},
  author={Wang, Jing and Lian, Biao and Zhang, Shou-Cheng},
  journal={Physica Scripta},
  volume={2015},
  number={T164},
  pages={014003},
  year={2015},
  publisher={IOP Publishing}
}

@article{Skinner2013Theory,
    author = {Skinner, Brian and Shklovskii, B. I.},
    title = {Theory of the random potential and conductivity at the surface of a topological insulator},
    journal = {Physical Review B},
    year = {2013},
    Volume = {87},
    number = {7},
    pages = {075454},
    publisher = {APS},
}

@article{drath1967band,
  title={Band parameters and g-factor for n-type {Bi$_2$Te$_3$} from the {Schubnikow-De Haas} effect},
  author={Drath, P and Landwehr, G},
  journal={Physics Letters A},
  volume={24},
  number={10},
  pages={504--506},
  year={1967},
  publisher={Elsevier}
}

@article{Mahmoodian2020Conductivity,
    author = {Mahmoodian, M. M. and Entin, M. V.},
    title = {Conductivity of a two-dimensional {HgTe} layer near the critical width: {The} role of developed edge states network and random mixture of p- and n-domains},
    journal = {Physical Review B},
    year = {2020},    
    volume = {101},
    number = {12},
    pages = {125415},
    publisher = {APS}
}

@article{neupane2014observation,
  title={Observation of quantum-tunnelling-modulated spin texture in ultrathin topological insulator {Bi$_2$Se$_3$} films},
  author={Neupane, Madhab and Richardella, Anthony and S{\'a}nchez-Barriga, Jaime and Xu, SuYang and Alidoust, Nasser and Belopolski, Ilya and Liu, Chang and Bian, Guang and Zhang, Duming and Marchenko, Dmitry and Varykhalov, Andrei and Rader, Oliver and Leandersson, Mats and Balasubramanian, Thiagarajan and Chang, Tay-Rong and Jeng, Horng-Tay and Basak, Susmita and Lin, Hsin and Bansil, Arun and Samarth, Nitin and Hasan, M. Zahid},
  journal={Nature communications},
  volume={5},
  number={1},
  pages={3841},
  year={2014},
  publisher={Nature Publishing Group UK London}
}

@article{kou2017two,
    author = {Liangzhi Kou and Yandong Ma and Ziqi Sun and Thomas Heine and Changfeng Chen},
    title = {Two-Dimensional Topological Insulators: {Progress} and Prospects},
    journal = {The Journal of Physical Chemistry Letters},
    year = {2017},
    volume = {8},
    number = {8},
    pages = {1905-1919},
    publisher = {ACS Publications}
}

@misc{supp,
    note = {Akin to a magnetization gap ($\Delta = g\mu_B B$), hybridization adds a mass term $M$ to the Dirac dispersion, opening a band gap in the surface state dispersion around $k=0$, the Dirac point (DP),
\begin{equation}\label{eq:gapDeltaM}
    E_\mathbf{k} = \pm \sqrt{v_F k^2 + (M -\mathcal{B}k^2 \pm \Delta)^2},
\end{equation}
with $v_F$ the Fermi velocity, and $\mathcal{B}$ a material dependent BHZ tight binding parameter. The hybridization gap is expected to close linearly in magnetic field.}

}

@article{he2012highly,
  title={Highly tunable electron transport in epitaxial topological insulator {(Bi$_{1-x}$Sb$_x$)$_2$Te$_3$} thin films},
  author={He, Xiaoyue and Guan, Tong and Wang, Xiuxia and Feng, Baojie and Cheng, Peng and Chen, Lan and Li, Yongqing and Wu, Kehui},
  journal={Applied Physics Letters},
  volume={101},
  number={12},
  year={2012},
  publisher={AIP Publishing}
}

@article{zhang2011growth,
  title={Growth of topological insulator Bi2Se3 thin films on SrTiO3 with large tunability in chemical potential},
  author={Zhang, Guanhua and Qin, Huajun and Chen, Jun and He, Xiaoyue and Lu, Li and Li, Yongqing and Wu, Kehui},
  journal={Advanced Functional Materials},
  volume={21},
  number={12},
  pages={2351--2355},
  year={2011},
  publisher={Wiley Online Library}
}

@article{chen2011tunable,
  title={Tunable surface conductivity in Bi 2 Se 3 revealed in diffusive electron transport},
  author={Chen, J and He, XY and Wu, KH and Ji, ZQ and Lu, L and Shi, JR and Smet, JH and Li, YQ},
  journal={Physical Review B—Condensed Matter and Materials Physics},
  volume={83},
  number={24},
  pages={241304},
  year={2011},
  publisher={APS}
}

@article{konig2013spatially,
  title={Spatially resolved study of backscattering in the quantum spin {Hall} state},
  author={K{\"o}nig, Markus and Baenninger, Matthias and Garcia, Andrei GF and Harjee, Nahid and Pruitt, Beth L and Ames, C and Leubner, Philipp and Br{\"u}ne, Christoph and Buhmann, Hartmut and Molenkamp, Laurens W and Goldhaber-Gordon, David},
  journal={Physical Review X},
  volume={3},
  number={2},
  pages={021003},
  year={2013},
  publisher={APS}
}

@article{li2017interaction,
  title={Interaction between counter-propagating quantum {Hall} edge channels in the {3D} topological insulator {BiSbTeSe$_2$}},
  author={Li, Chuan and De Ronde, Bob and Nikitin, Artem and Huang, Yingkai and Golden, Mark S and De Visser, Anne and Brinkman, Alexander},
  journal={Physical Review B},
  volume={96},
  number={19},
  pages={195427},
  year={2017},
  publisher={APS}
}

@article{fatemi2014electrostatic,
  title={Electrostatic coupling between two surfaces of a topological insulator nanodevice},
  author={Fatemi, Valla and Hunt, Benjamin and Steinberg, Hadar and Eltinge, Stephen L and Mahmood, Fahad and Butch, Nicholas P and Watanabe, Kenji and Taniguchi, Takashi and Gedik, Nuh and Ashoori, Raymond C and Jarillo-Herrero, Pablo},
  journal={Physical review letters},
  volume={113},
  number={20},
  pages={206801},
  year={2014},
  publisher={APS}
}

@article{chong2019tunable,
  title={Tunable coupling between surface states of a three-dimensional topological insulator in the quantum {Hall} regime},
  author={Chong, Su Kong and Han, Kyu Bum and Sparks, Taylor D and Deshpande, Vikram V},
  journal={Physical Review Letters},
  volume={123},
  number={3},
  pages={036804},
  year={2019},
  publisher={APS}
}

@article{yang2015dual,
  title={Dual-gated topological insulator thin-film device for efficient fermi-level tuning},
  author={Yang, Fan and Taskin, AA and Sasaki, Satoshi and Segawa, Kouji and Ohno, Yasuhide and Matsumoto, Kazuhiko and Ando, Yoichi},
  journal={Acs Nano},
  volume={9},
  number={4},
  pages={4050--4055},
  year={2015},
  publisher={ACS Publications}
}

@article{lu2011competition,
  title={Competition between weak localization and antilocalization in topological surface states},
  author={Lu, Hai-Zhou and Shi, Junren and Shen, Shun-Qing},
  journal={Physical review letters},
  volume={107},
  number={7},
  pages={076801},
  year={2011},
  publisher={APS}
}

@article{kolling2025gate,
  title={Gate electrode-induced nonreciprocal resistance in topological insulators},
  author={K{\"o}lling, Sofie and Westerhof, Florian R and Brinkman, Alexander},
  journal={arXiv preprint arXiv:2503.13141},
  year={2025}
}

@misc{refsupp,
  howpublished = {See Supplemental Material for magnetoresistance and additional differential conductance data. See also references [43-51] therein.}
}

@article{jiang2012landau,
  title={Landau quantization and the thickness limit of topological insulator thin films of {Sb$_2$Te$_3$}},
  author={Jiang, Yeping and Wang, Yilin and Chen, Mu and Li, Zhi and Song, Canli and He, Ke and Wang, Lili and Chen, Xi and Ma, Xucun and Xue, Qi-Kun},
  journal={Physical review letters},
  volume={108},
  number={1},
  pages={016401},
  year={2012},
  publisher={APS}
}

@article{kolling2025non,
  title={{Non-Ohmic behavior in (Bi $ \_ $\{$1-x$\}$ $ Sb $ \_x $) $ \_2 $ Te $ \_3 $ by Joule heating}},
  author={K{\"o}lling, Sofie and Wielens, Daan H and Adagideli, {\.I}nan{\c{c}} and Brinkman, Alexander},
  journal={arXiv preprint arXiv:2505.11138},
  year={2025}
}

@article{anderson1979possible,
  title={Possible explanation of nonlinear conductivity in thin-film metal wires},
  author={Anderson, PW and Abrahams, E and Ramakrishnan, TV},
  journal={Phys. Rev. Lett.},
  volume={43},
  number={10},
  pages={718},
  year={1979},
  publisher={APS}
}

\end{document}


\maketitle
\supplementarysection
\vfill
\tableofcontents{}
\vfill

\clearpage
\FloatBarrier
\section{HLN fit of weak anti-localization}

\begin{figure}[h!]
    \centering
    \includegraphics[width = \linewidth]{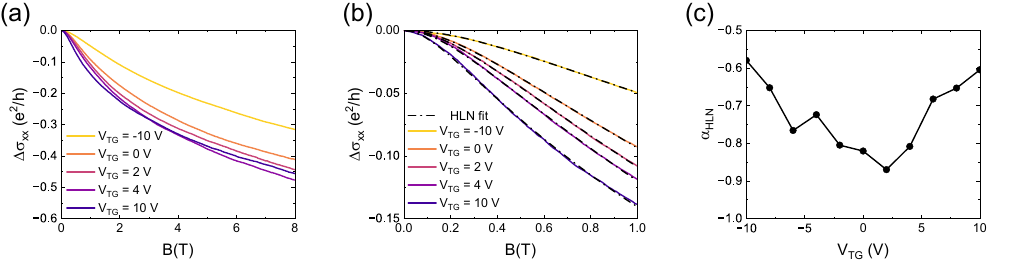}
    \caption{Weak anti-localization (WAL) in the sample corresponding to Fig. 3 in the main text, measured at $T$ = 4.5 K. (a) Conductivity correction ($\Delta\sigma_{xx}$) for $V_\mathrm{TG}$ = -10 V, 0 V, 2 V, 4 V and 10 V. (b) dataset from (a), zoomed in and including fits of the Hikami-Larkin-Nagaoka (HLN) equation. (c) $\alpha_\mathrm{HLN}$ obtained by fitting the HLN equation as shown in (b), versus gate voltages ranging from $V_\mathrm{TG}$ = -10 V up to $V_\mathrm{TG}$ = 10 V in steps of $\Delta V$ = 2V. For clarity, not all curves for which the $\alpha_\mathrm{HLN}$ parameter is obtained are plotted in (a), (b) and Fig. 3 in the main text.}
    \label{fig:suppl_WAL}
\end{figure}

Figure~\ref{fig:suppl_WAL}(a) shows the symmetrized correction to the longitudinal conductivity showing clear signatures of WAL. In Fig.~\ref{fig:suppl_WAL}(b), the Hikami-Larkin-Nagaoka equation is fitted using the rage between $B$ = 0 T and $B$ = 1 T \cite{hikami1980spin}, using HLN parameter $\alpha_\mathrm{HLN}$ and phase coherence length $L_\phi$ as fit parameters. It can be seen in the figure that the HLN-equation fits reasonably well. 
The fit parameter $\alpha_\mathrm{HLN}$ is shown in Fig.~\ref{fig:suppl_WAL}(c). In a similar material it was shown that in a TI where two surfaces are decoupled (coupled), $\alpha_\mathrm{HLN}$ = -1 (-0.5) \cite{steinberg2011electrically}. Here, the system is observed to be in an intermediate regime with a value of $\alpha_\mathrm{HLN} \approx$ 0.75. Numerous explanations can be found for such values of $\alpha_\mathrm{HLN}$, such as surface-to-bulk coupling \cite{steinberg2011electrically}, a gap opening at the Dirac point \cite{Lang2012Competing, lu2011competition} or a mix of these effects \cite{Lu2011Weak}, and therefore, we stress that the interpretation of this $\alpha_\mathrm{HLN}$ parameter is ambiguous.

\FloatBarrier
\clearpage
\section{VI characteristics at elevated temperatures}

\begin{figure}[h!]
    \centering
    \includegraphics[width=0.5\linewidth]{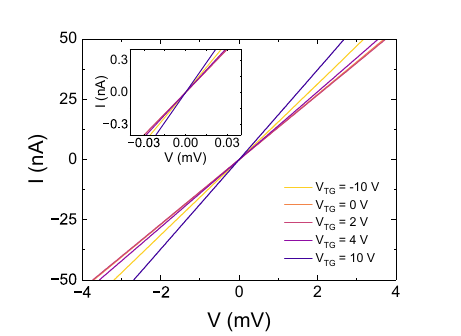}
    \caption{IV characteristics of the device of Fig. 3 in the main text, measured at $T$ = 4.5 K and for $V_\mathrm{TG}$ = -10 V, 0 V, 2 V, 4 V and 10 V. The inset shows the same measurement zoomed in around $I$ = 0 nA.}
    \label{fig:suppl_IV}
\end{figure}

In Fig. 3 in the main text, the gate-dependent magnetotransport characteristics are analyzed at elevated temperatures. Figure \ref{fig:suppl_IV} shows the IV characteristics corresponding to the device measured in Fig. 3. While this measurement is performed using a current source, here, the axis are interchanged for better comparison with the measurement shown later in \ref{fig:suppl_dIdVMaps}(c-d). As can be seen in figure \ref{fig:suppl_IV} no signatures of hybridization are found. At low temperatures however, signatures appear in this device, but the dataset is less conclusive as a current bias was used.

\newpage
\section{Geometry and temperature dependence of gap}

\begin{figure}[h!]
    \centering
    \includegraphics{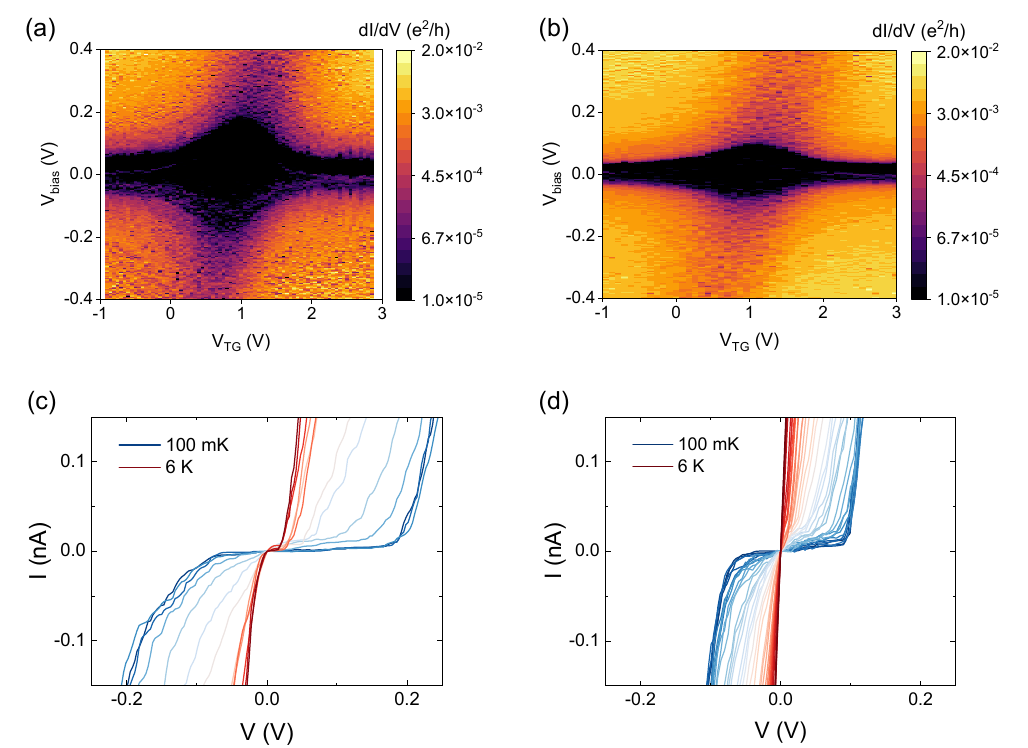}
    \caption{(a-b) Differential conductance maps from two Hall bars with Au contacts at $T = 100$ mK. (a) $L\times W = 80$ $\mu$m$\times 5$ $\mu$m, of which the 4-terminal differential conductance map is shown in Fig. 4(b) in the main text. (b) $L\times W = 37.5$ $\mu$m$\times 5$ $\mu$m. Both (a) and (b) are obtained in a 2-terminal setup, and both show skewed-ness likely due to the interplay between bias and gate voltage. The gap in (a) is enhanced with respect to the gap in (b). (c-d) IV-curves from the devices in (a-b) while $V_\mathrm{TG} = 1$ V, between $T = 100$ mK and 6 K. The temperatures correspond to the data points in Fig.~\ref{fig:disorder_electr}(b)}
    \label{fig:suppl_dIdVMaps}
\end{figure}

Having found that ultrathin TIs host an insulating gap that is intrinsic to the film in the main text, we elaborate on the analysis by studying the temperature- and geometry-dependence of this gap.
Figures~\ref{fig:suppl_dIdVMaps}(a-b) show the differential conductance maps from two devices with Au contacts in a 2-terminal setup at $T = 100$ mK. The corresponding IV curves extracted at $V_\mathrm{TG} = 1$ V are shown in Fig.~\ref{fig:suppl_dIdVMaps}(c-d) for a range of temperatures. The gap is visible as a flat (insulating) region around $V = 0$ V.
The asymmetry in the IV curve for positive and negative gate voltages can be related to the interplay between bias and gate voltage, also visible as the skewed behavior in Fig. 4(b). Correcting for this mechanism is not straightforward partially due to the variable resistance as function of bias, reminiscent of a dual-gated topological insulator \cite{fatemi2014electrostatic, chong2019tunable, yang2015dual}. Therefore, in further analysis we approximate this interplay as linear and consider the averaged behavior between positive and negative bias.

\begin{figure*}[h!]
    \centering\includegraphics[width=\textwidth]{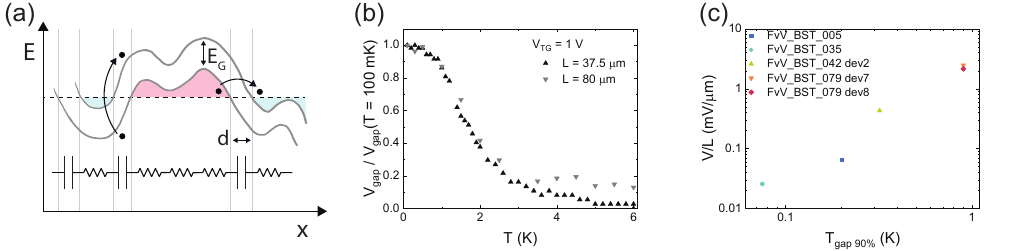}
    \caption{(a) Disordered potential landscape in an ultrathin topological insulator. Charge puddles vary the position of the hybridization gap (size $E_\mathrm{G}$) compared to the Fermi level, creating electron (blue) and hole (red) pockets. Conduction is enabled by thermal excitation over the gap, or by tunneling through the insulating regimes, of characteristic thickness $d$. In the DC limit, an analogy can be drawn with a series of resistive and capacitive circuit elements as illustrated along the $x$-axis. Image adapted from \cite{nandi2018signatures}. (b) Extracted threshold voltage $V_\mathrm{gap}$ from Fig.~\ref{fig:suppl_dIdVMaps}(c-d) normalized to the $T = 100$ mK values. The temperature dependence of the normalized gap is similar in both devices. (c) Hybridization gap in devices fabricated from different ultrathin BST films (distinguished by symbol shape). The voltage gradient required to close the gap ($V/L$) at $T = 100$ mK is shown versus the temperature at which $V_\mathrm{gap}$ is at 90\% of its 100 mK value to compare the datasets.}
    \label{fig:disorder_electr}
\end{figure*}

When $V$ exceeds the threshold value, the device becomes conducting. We find that the average threshold value for positive and negative bias, $V_\mathrm{gap}$, is higher for the longer device in Fig. \ref{fig:suppl_dIdVMaps}(c) than for the shorter device in Fig. \ref{fig:suppl_dIdVMaps}(d). However, when normalizing $V_\mathrm{gap}$ extracted from the IV curves to the $T = 100$ mK value in Fig. \ref{fig:disorder_electr}(b), both devices have a similar temperature dependence.

An explanation for the geometry-induced variation in $V_\mathrm{gap}$ might lie in the disordered nature of our BST thin films. Although BST possesses properties beneficial for a transport study, dopants are also known to affect material properties significantly \cite{Skinner2013Theory}. In particular, if a (hybridization) band gap is probed by performing a transport study, often a smaller gap is observed than the intrinsic band gap of the system due to the presence of Coulomb impurities \cite{Huang2022conductivity, knispel2017charge}. In other words, the hybridization gap can fluctuate in size and position with respect to the Fermi level.

We use a simplified model to explain the temperature- and geometry-dependence of the observed insulating regime. The potential landscape of the disordered ultrathin topological insulator consists of insulating and metallic regions around the Fermi level, as illustrated in Fig.~\ref{fig:disorder_electr}(a). This system can be compared to a circuit of capacitive and resistive elements connected in series. When applying a potential gradient to the disordered potential landscape, conduction is enabled when the width of the tunnel barrier between electron and hole puddles, which an in-plane electric field can reduce, falls below a threshold value \cite{nandi2018signatures}. In the electric circuit analogy, this effect can be captured as a finite voltage rating of the capacitive elements, above which the dielectric breaks down (note that we only consider this breakdown-behavior, and do not extend the analogy to the time-dependent charging and discharging of the capacitive elements). When increasing the device length, \textit{e.g.} increasing the number of capacitive elements in series, the voltage rating of the complete circuit is reduced. In other words, due to the disordered potential landscape, the threshold voltage required for conduction scales with the device length.

However, conduction by thermal excitation in this system depends only on the gap size and is independent of the distance between charge puddles. Considering this, we expect no difference in the temperature at which the insulating regime vanishes between two devices with a different geometry (fabricated from the same ultrathin film). This is in line with the data in Fig. \ref{fig:disorder_electr}(b): the temperature dependence of the gap is independent of geometry, whereas the voltage required to cross the gap scales with device length.

The above considerations show that the hybridization gap cannot be directly obtained by only studying the bias voltage at which a gap is closed. 
Device geometry is also known to affect transport in 2D QSH systems that are near the topological transition point \cite{Mahmoodian2020Conductivity}, however, to our knowledge the role of device geometry on measurable activation energy remains unknown.

We show the reproducibility of the observed gapped region by extending this analysis to multiple devices fabricated from a range of ultrathin BST films. Despite the fact that all films are deposited with the same deposition parameters, the impurity density and hybridization gap might vary due to inherent film-to-film variation in the MBE deposition. We cannot obtain the gap from solely the threshold voltage, as mentioned above. However, we can obtain information about the nature of the gap by comparing the threshold voltage gradient (electric field) to the temperature at which the insulating regime vanishes, since both quantities are proportional to the gap size.

When comparing $V_\mathrm{gap}$ extracted from IV curves (normalized by device length) to the critical temperature at which $V_\mathrm{gap}(T) = 0.9\cdot V_\mathrm{gap}(100\ \mathrm{mK})$ (a measure to compare the temperature dependence of the gap between different datasets), a general scaling emerges in Fig.~\ref{fig:disorder_electr}(c). Although $V_\mathrm{gap}$ fluctuates between devices, the gradient $V_\mathrm{gap}/L$ (dependent on device geometry) and critical temperature (independent of geometry) are correlated. This general trend shows that a similar mechanism underlies the insulating behavior in different films. We attribute this to a reproducible hybridization gap in ultrathin BST, although the potential landscape varies film-to-film.
\newpage
\FloatBarrier

\printbibliography
\FloatBarrier